\documentclass[aps,prl,groupedaddress,showpacs]{revtex4}

\usepackage{amsfonts}
\usepackage{graphicx}

\begin{document}


\title{Efficient Plasma Heating by Radiofrequency}


\author{Marco Frasca}
\email[]{marcofrasca@mclink.it}
\affiliation{Via Erasmo Gattamelata, 3 \\ 00176 Roma (Italy)}


\date{\today}

\begin{abstract}
We propose an efficient mechanism to heat a plasma by an intense microwave field solving the equation of
ion motion in a wave field and a constant magnetic field in a large coupling regime. The 
mechanism does not relies explicitly on stochastic motion and is able to increase the
ion velocity by several magnitude orders. Known thresholds for the onset of stochastic motion
are also obtained.
\end{abstract}


\maketitle


The possibility to realize confined nuclear fusion is linked to the ability to let ions
reach enough high energies to ignite the process. Heating of
such plasmas can be obtained by different means the most
common of which is by radiofrequency directly injected through the
confining system that is generally a tokamak.

Such a heating exploits directly the interaction between a charged particle
and an electromagnetic mode of a field propagating inside the magnetically
confined plasma. This kind of interaction, being ruled by a strongly non-linear
equation, is rather involved and a lot of analysis have been performed in the
latest thirty years to properly understand the process.

The most exploited effect that happens in such wave-particle interaction is
the onset of a stochastic behavior according to Kolmogorov-Arnold-Moser
(KAM) theorem \cite{sk,ka1,ka2,ka3,sat}. The availability of a large portion of phase
space to the ions, when a threshold is overcome, and the consequent
entering into a higher energy regime is termed ``stochastic heating'' and it
proves to be a very efficient method to heat a plasma by radiofrequency.

Stochastic heating is commonly analyzed through a well-known model that, with the
proper modifications, is able to give an adequate understanding of the physics
at hand. Indeed, for an ion with charge $q$ and mass $m$ with a constant
magnetic field ${\bf B}=(0,0,B_0)$ and a wave ${\bf E}=(0,E_0,0)\cos(ky-\omega t)$,
one has the equation\cite{sk,ka1,ka2,ka3}
\begin{equation}
     \ddot y(t)+\Omega^2 y(t)=\frac{qE_0}{m_0}\cos(ky(t)-\omega t)
\end{equation}
being $\Omega=qB_0/m_0$ the cyclotron frequency. We make this equation adimensional by
introducing $y(t)\rightarrow ky(t)$, $t'=\Omega t$, $\alpha=E_0k/B_0\Omega$ and $\nu=\omega/\Omega$
so that, with the above redefinition of $y$,
\begin{equation}
     \ddot y(t')+y(t')=\alpha\cos(y(t')-\nu t')
\end{equation}
where we recognize that two critical parameters fix the physics for this problem, $\alpha$
and $\nu$. This is a Hamiltonian system.
Indeed, it known that for stochastic heating produced by a lower hybrid wave
the stochasticity threshold, assuming $\nu>1$, is given by \cite{ka1,ka2,ka3} 
\begin{equation}
\label{eq:th}
    \alpha\approx\frac{1}{4}\nu^{\frac{2}{3}}
\end{equation}
while for ion cyclotron heating this condition becomes \cite{sk} $\alpha\approx 1$. Dissipation of the
energy of the wave is due to Landau damping that, for the presence of the magnetic field,
makes the absorption irreversible \cite{suz}.

The approach that is normally adopted to cope with this kind of problems is to work out canonical
perturbation theory that holds for small $\alpha$. Then, Chirikov overlap criterion for resonances \cite{chi} 
and comparison with numerical results are used
to determine the threshold of the onset of stochasticity to see when effective heating starts to set in.
In this situation physics is fairly well-known and a diffusion equation can be derived describing
the behavior of the plasma under such conditions. The overall mark of stochastic heating mechanism
is the fact that the system under study is not integrable and KAM theorem does apply.

Anyhow, we note that the opposite limit $\alpha\rightarrow\infty$ gives us the interesting result that
the system is ruled by the equation 
\begin{equation}
    \ddot y_0(t')=\alpha\cos(y_0(t')-\nu t')
\end{equation}
that is an integrable one. Indeed, the solution to this equation is straightforward to write down as 
\begin{equation}
\label{eq:exa}
   y_0(t')=\frac{\pi}{2}+\nu t'+2\arcsin(m{\rm sn}(\sqrt{\alpha}t',m))
\end{equation}
being $\rm sn$ the Jacobi snoidal elliptic function and $m$ its modulus. We notice that if 
we take $y(0)=\pi/2$ and $\dot y(0)=kv_0/\Omega$ it easy to see that 
\begin{equation}
    m=\frac{1}{2\sqrt{\alpha}}\left(\frac{kv_0}{\Omega}-\nu\right)
\end{equation}
and then the modulus of the Jacobi function depends on the initial
velocity of the ion. Besides, $T_{tr}=4K(m)/\sqrt{\alpha}$, with 
$K(m)=\int_0^{\frac{\pi}{2}}d\phi/\sqrt{1-m^2\sin\phi}$, is the bounce time of a ion trapped
inside a well of the wave. From this exact solution we can recognize two regimes. 
One has the ions moving uniformly with the phase velocity of the wave
and the other is just the contribution of the trapped motion. If the first effect
prevails one can have a net increase of the energy of the ion that, for the presence of the magnetic field,
is irreversibly absorbed. Indeed, this can be assured by taking $m\ll 1$ so that one can expand the snoidal
function to have
\begin{equation}
\label{eq:app}
    y_0(t')\approx\frac{\pi}{2}+\nu t'+2m\sin(\sqrt{\alpha} t') + O(m^2) 
\end{equation}
and an efficient heating
mechanism is granted. In fact, the averaged velocity on a bounce time will be given by 
$\langle\dot y_0\rangle=\nu$. This means that the
ion is moving with the phase velocity of the wave that is assumed much larger of the initial ion velocity. 
Similarly, one has
$\langle\dot y_0^2\rangle=\nu^2$ and a diffusion coefficient can be estimated dividing by
the bounce time as $D=\langle\dot y_0^2\rangle/2T_{tr}\approx \pi\nu^2/\sqrt{\alpha}$.

This scenario would be consistent if we would be able to develop an asymptotic perturbation series
in the limit $\alpha\rightarrow\infty$ to compute higher order corrections due to the presence of the
magnetic field. Indeed this can be accomplished by the duality principle in perturbation theory 
\cite{fra1,fra2,fra3,fra4,fra5}. This principle can be stated by saying that interchanging the perturbation
terms in an equation gives two perturbation series having a development parameter that in a series
is the inverse of the other.
That is, in our case one would have (weak perturbation, $\alpha\rightarrow 0$) 
$y=y_0+\alpha y_1+\alpha^2 y_2+\ldots$ and
for the dual series $y=y_0+\alpha^{-1} y_1+\alpha^{-2} y_2+\ldots$ 
that holds in the opposite limit $\alpha\rightarrow\infty$.
In order to get a non trivial set of perturbation equations, one need to rescale time \cite{fra1,fra2} because,
as already seen above, this is the proper scaling of the leading order solution. So, we take
\begin{eqnarray}
    \tau &=& \sqrt{\alpha}t' \\ \nonumber
    y_0 &=& y_0+\alpha^{-1} y_1+\alpha^{-2} y_2+\ldots
\end{eqnarray}
into the motion equation and the following working set of equations is easily obtained
\begin{eqnarray}
    \ddot y_0(\tau)-\cos\left(y_0(\tau)-\frac{\nu}{\sqrt{\alpha}}\tau\right) &=& 0 \\ \nonumber
    \ddot y_1(\tau)+\sin\left(y_0(\tau)-\frac{\nu}{\sqrt{\alpha}}\tau\right)y_1(\tau) &=& -y_0(\tau) \\ \nonumber
    \ddot y_2(\tau)+\sin\left(y_0(\tau)-\frac{\nu}{\sqrt{\alpha}}\tau\right)y_2(\tau) &=& -y_1(\tau) 
    -\frac{1}{2}\cos\left(y_0(\tau)-\frac{\nu}{\sqrt{\alpha}}\tau\right)y_1^2(\tau) \\ \nonumber
    &\vdots&
\end{eqnarray}
Using the exact solution to the leading order (\ref{eq:exa}) the above set of equations takes the form
\begin{eqnarray}
    \ddot y_1(\tau)+[1-2m^2{\rm sn}^2(\tau,m)]y_1 &=& -y_0(\tau) \\ \nonumber
    \ddot y_2(\tau)+[1-2m^2{\rm sn}^2(\tau,m)]y_2 &=& -y_1(\tau) 
    +m{\rm sn}(\tau,m)\sqrt{1-m^2{\rm sn}^2(\tau,m)}y_1^2(\tau) \\ \nonumber
    &\vdots&
\end{eqnarray}
that in the limit $m\ll 1$ take the simpler form
\begin{eqnarray}
    \ddot y_1(\tau)+ y_1 &=& -y_0(\tau) \\ \nonumber
    \ddot y_2(\tau)+ y_2 &=& -y_1(\tau) 
    +m\sin(\tau)y_1^2(\tau) \\ \nonumber
    &\vdots&
\end{eqnarray}
to the first order in $m$. These equations can be solved straightforwardly taking at the
leading order eq.(\ref{eq:app}).

The solutions contain secular terms. These terms can be removed in many ways. The one we prefer
is the renormalization group method as presented in \cite{rg1,rg2,rg3}. This technique
assumes that the perturbation solution should be computed to an initial time $\tau_0$ as to
have $y(\tau,\tau_0)=y_0(\tau,\tau_0)+\alpha^{-1} y_1(\tau,\tau_0)+\alpha^{-2} y_2(\tau,\tau_0)+\ldots$
and the initial constants depend on the initial time, $\phi=\phi(\tau_0)$ and $m=m(\tau_0)$. The
requirement that the perturbation series should not depend on the choice of the initial point translates
into the condition
\begin{equation}
     \left.\frac{dy(\tau,\tau_0)}{d\tau_0}\right|_{\tau_0=\tau}=0
\end{equation}
that gives back renormalization group equations for the constants $\phi$ and $m$ that now turn out to depend on 
$\tau$. Then, the condition $\left.y(\tau,\tau_0)\right|_{\tau_0=\tau}$ produces the correct 
perturbation series without secular terms. It is interesting to point out that all we have done is just to
compute an envelope \cite{rg3}.

After resummation we get
\begin{eqnarray}
    y_0(\tau)&=&\frac{\pi}{2}+\frac{\nu}{\sqrt{\alpha}}\tau+2m(\tau)\sin(\tau+\phi(\tau)) \\ \nonumber
     y_1(\tau) &=& 
     \frac{\pi}{2}[\cos(\tau+\phi(\tau))-1]-m(\tau)\sin(\tau+\phi(\tau))) \\ \nonumber
     y2(\tau) &=& \frac{\pi}{2}[1-\cos(\tau+\phi(\tau))]
     +\left(\frac{\pi^2}{48}+\frac{3}{4}\right)m(\tau)\sin(\tau+\phi(\tau)) \\ \nonumber
     & &-\frac{\pi^2}{32}m(\tau)\sin(\tau+\phi(\tau))\cos^2(\tau+\phi(\tau))
     +\frac{\pi^2}{6}m(\tau)\sin(\tau+\phi(\tau))\cos(\tau+\phi(\tau))
\end{eqnarray}
where now the modulus $m$ depends on $\tau$ and the phase $\phi(\tau)$ contains a frequency shift as from the
renormalization group equations
\begin{eqnarray}
    \frac{dm}{d\tau} &=& -\frac{\pi}{8\alpha^2}+O(\alpha^{-3}) \\ \nonumber
    \frac{d\phi}{d\tau} &=& \frac{1}{2\alpha}
    -\frac{1}{4\alpha^2}\left(\frac{5\pi^2}{8}+3\right)+O(\alpha^{-3}).
\end{eqnarray}
Having the modulus depending linearly on $\tau$ implies that if one waits a time long enough our initial approximation
$m\ll 1$ in absolute value may fail. We discuss this point below about stochasticity thresholds. We have also found
higher order corrections to the bounce time. Finally, we note the renormalization constant 
$1-1/\alpha+1/\alpha^2+\ldots=\alpha/(1+\alpha)$ to the $\pi/2$ term.

We now discuss the stochasticity threshold in the case of a ion cyclotron wave (ICW) \cite{sk} and lower hybrid wave
(LHW)\cite{ka1,ka2}. The mechanism of heating does work only if the particle moves freely at the phase velocity
of the wave. Instead, if bouncing of the ion in a well of the wave prevails, heating is not effective.
For a ICW one has $\nu=1$ and the only guaranty of a fully consistency of the above discussion is $\alpha\gg 1$
in agreement with the result given in \cite{sk}. 
For a LHW, when $\alpha$ is not so large but greater than one and $\nu\gg 1$
as discussed in \cite{ka1,ka2}, it is fundamental that the modulus $m$ does not prevail on the free particle term
forcing the ion behavior to a simple bouncing in the wells of the wave. This is granted taking 
$1/\nu>>\pi/8\alpha^{\frac{3}{2}}$ giving us the improved threshold for this wave
\begin{equation}
    \alpha\gg\frac{(\pi\nu)^{\frac{2}{3}}}{4}
\end{equation}
and the numerical factor is now about $0.54$. In this way we have proved fully consistency of our approach
with respect to previous ones on stochastic heating obtaining a deeper understanding of the involved physics.

In order to complete our analysis we give here some numerical studies that are essential
to support the above scenario. We have considered two cases for $\alpha$, $\nu$ and $\dot y(0)$.
We have not taken $\alpha$ too large because otherwise the plots would not be much interesting
as the analytical solution hits the exact one as does the numerical one. The goodness of the
approximation we have applied is evident provided that $m$ is kept small and the thresholds
given above are sufficiently overcome. 
\begin{figure}[tbp]
\begin{center}
\includegraphics[width=240pt]{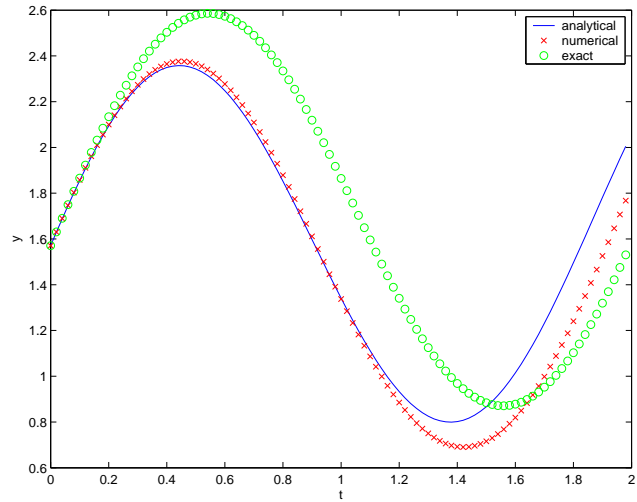}
\caption{\label{fig:fig1} Numerical, analytical and eq.(\ref{eq:exa})
for $\alpha=10$, $\nu=15$ and $\dot y(0)=0.01$.}
\end{center}
\end{figure}
So, fig.\ref{fig:fig1} corresponds to have $|m|\approx 0.45$ but $\alpha=10$
is not so large. 
\begin{figure}[tbp]
\begin{center}
\includegraphics[width=240pt]{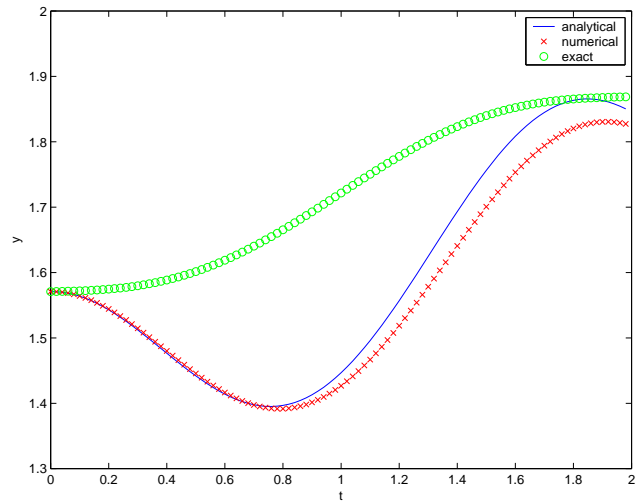}
\caption{\label{fig:fig2} Numerical, analytical and eq.(\ref{eq:exa})
for $\alpha=10$, $\nu=0.15$ and $\dot y(0)=0.01$..}
\end{center}
\end{figure}
The same happens in fig.\ref{fig:fig2} that has $|m|\approx 0.022$. 
\begin{figure}[tbp]
\begin{center}
\includegraphics[width=240pt]{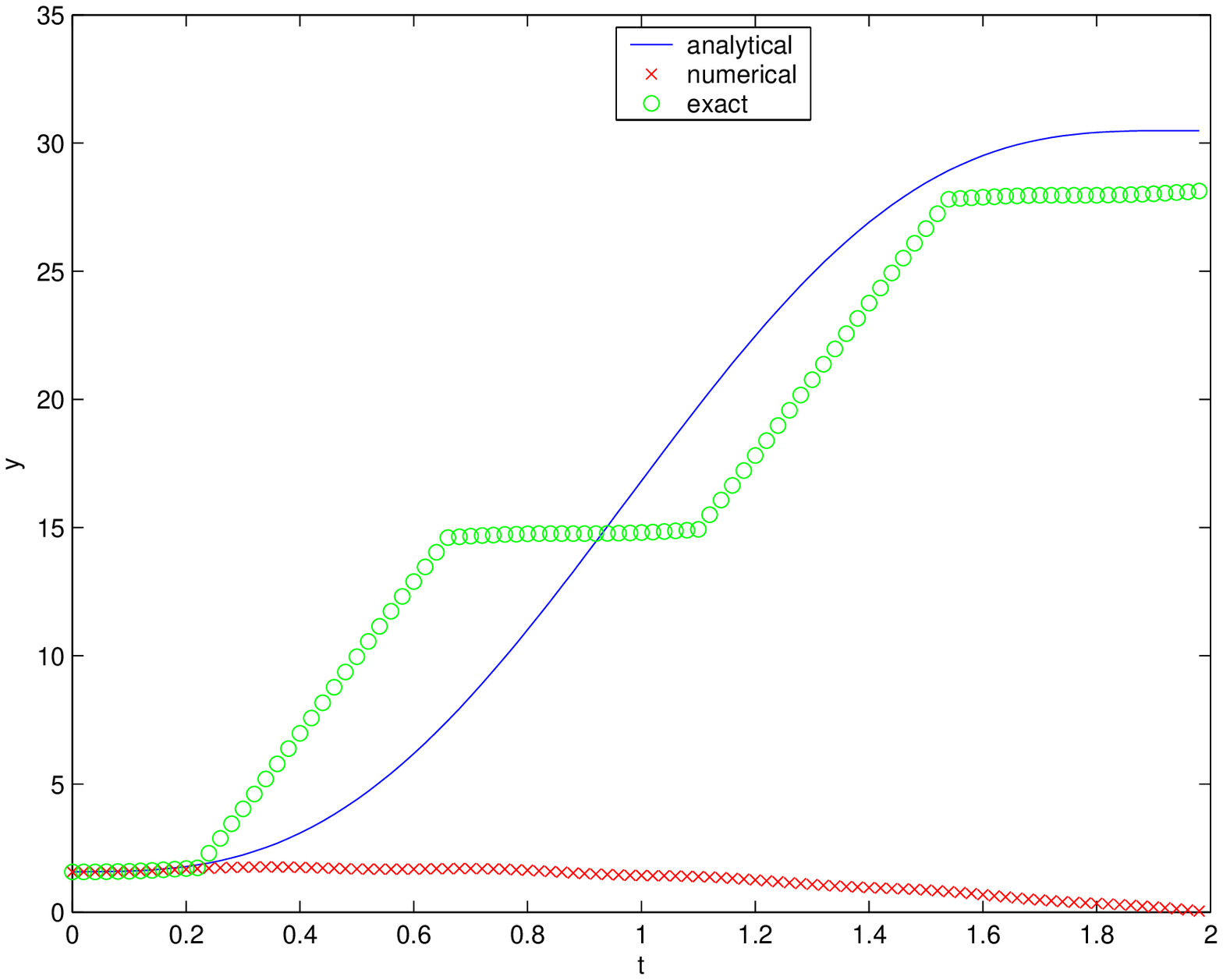}
\caption{\label{fig:fig3} Numerical, analytical and eq.(\ref{eq:exa})
for $\alpha=10$, $\nu=0.15$ and $\dot y(0)=3$.}
\end{center}
\end{figure}
The worst situation is seen in
fig.\ref{fig:fig3} where $|m|\approx 2.3$ and the thresholds not properly overcome. 
So due to the value of $\alpha=10$ and the corresponding
problems in $m$ and thresholds the agreement goes from relatively good to bad. The relevance
of the threshold is due both to the largeness of $\alpha$ and to avoid that $m$
increasing in time may overcome the initial condition $m\ll 1$ from which we started.
\begin{figure}[tbp]
\begin{center}
\includegraphics[width=240pt]{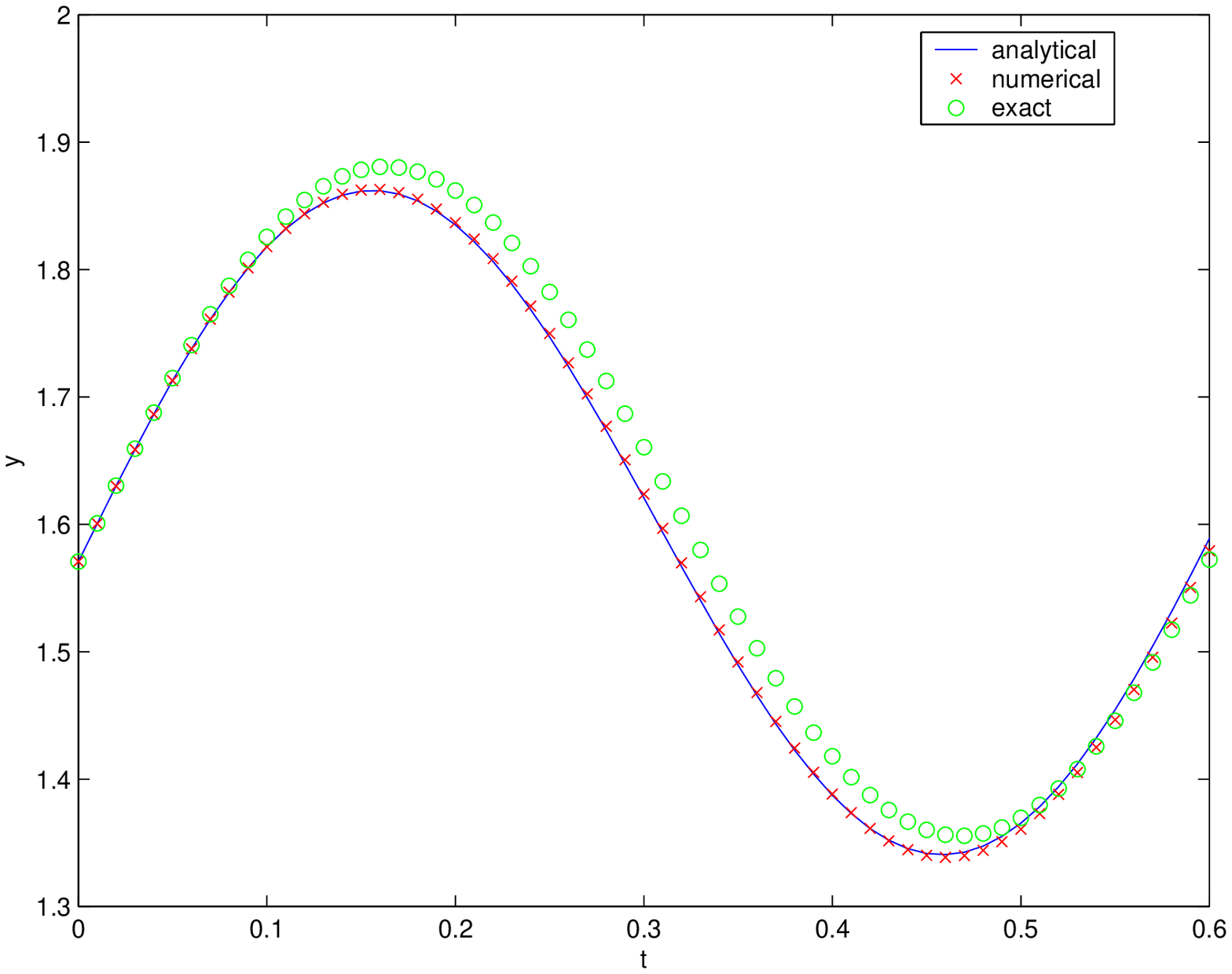}
\caption{\label{fig:fig4} Numerical, analytical and eq.(\ref{eq:exa})
for $\alpha=100$, $\nu=15$ and $\dot y(0)=0.01$.}
\end{center}
\end{figure}
Things improve,
as should be, with $\alpha=100$ where the agreement is very satisfactory. 
\begin{figure}[tbp]
\begin{center}
\includegraphics[width=240pt]{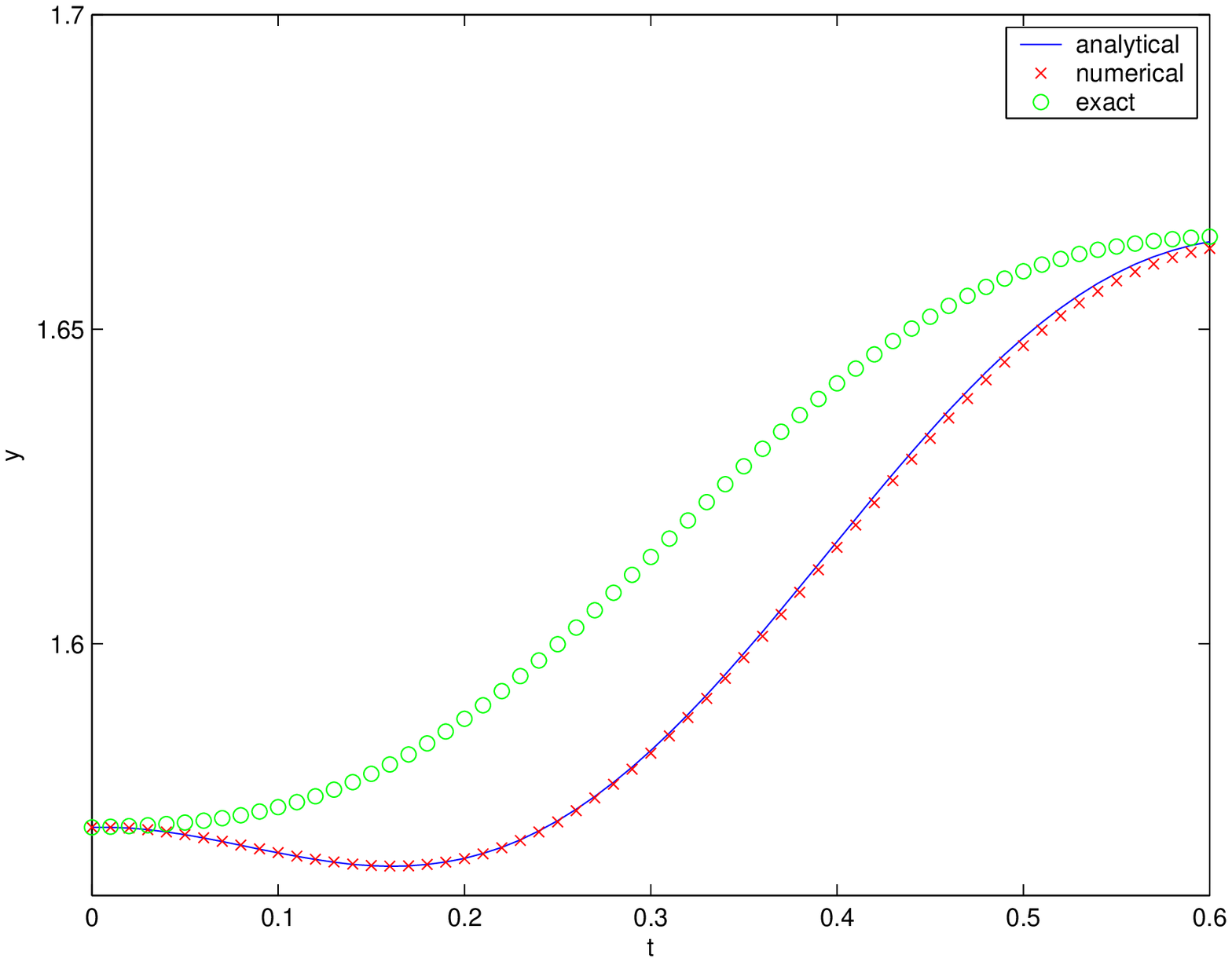}
\caption{\label{fig:fig5} Numerical, analytical and eq.(\ref{eq:exa})
for $\alpha=100$, $\nu=0.15$ and $\dot y(0)=0.01$.}
\end{center}
\end{figure}
\begin{figure}[tbp]
\begin{center}
\includegraphics[width=240pt]{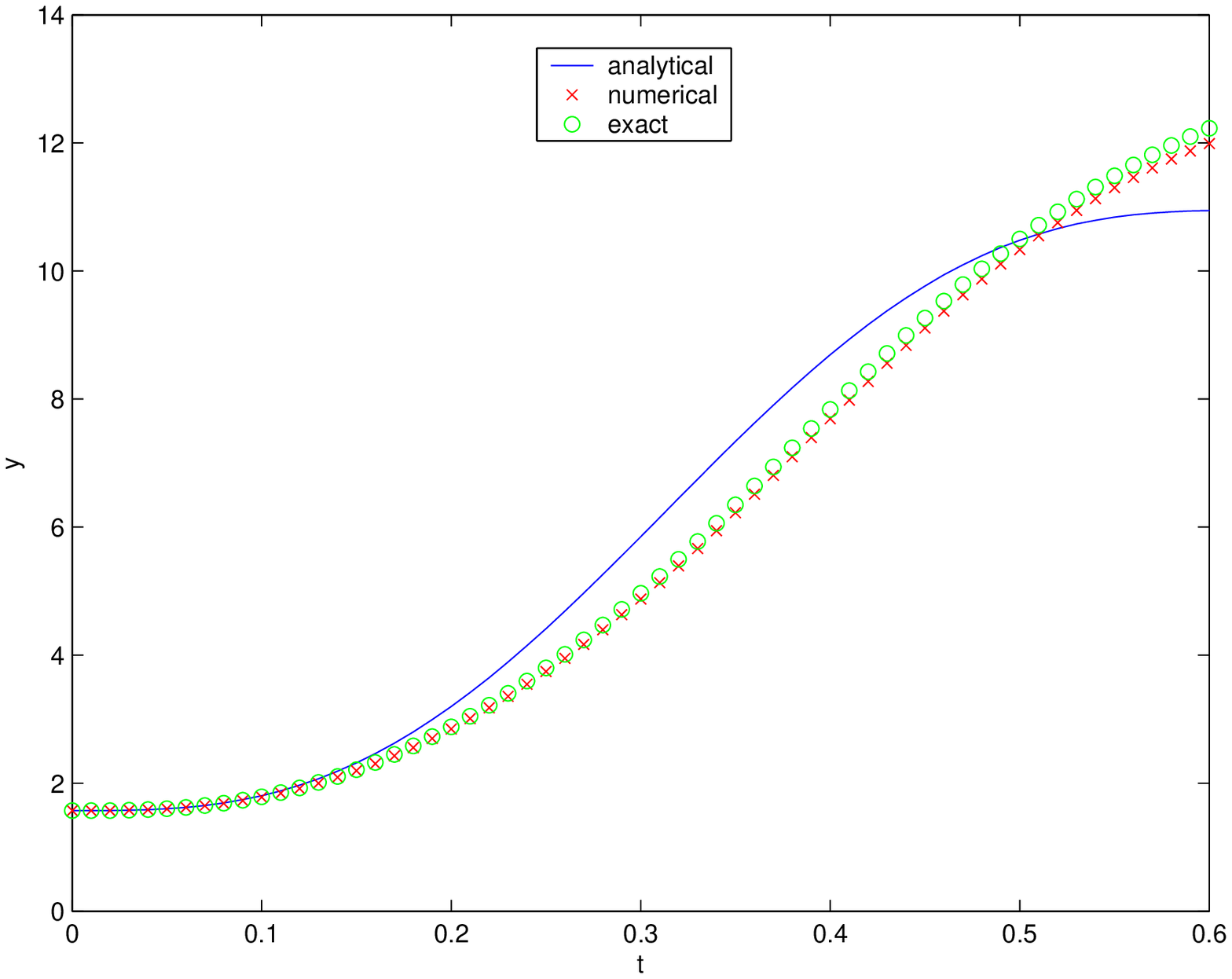}
\caption{\label{fig:fig6} Numerical, analytical and eq.(\ref{eq:exa})
for $\alpha=100$, $\nu=0.15$ and $\dot y(0)=3$.}
\end{center}
\end{figure}
An interesting
situation is seen in fig.\ref{fig:fig6} where the solution given in eq.(\ref{eq:exa}) is better
than the perturbation series. This is due to the fact that, although the thresholds are 
properly overcome, one has $m\approx 0.75$ very near one, where our fundamental assumption
to derive the perturbation series starts to fail. As expected, in some cases the perturbation
series performs much better than eq.(\ref{eq:exa}).

So, we can conclude that a satisfactory perturbation approach has been presented here to treat problem
of strong particle-wave interaction in a plasma. A heating scenario for ions has been
given showing that it can increase the ion velocity of several magnitude orders making
it move to the phase velocity of the wave. Known thresholds have been derived in a fully
analytical and rigorous way forming a complete scenario for plasma heating by radiofrequency.
A lot of problems having a longstanding numerical treatment could be faced now in an analytical
way by the method of dual perturbation theory giving an improved insight into physics.


\end{document}